\documentstyle[12pt]{article}
\newlength{\extraspace}
\newlength{\extraspaces}
\newcommand{\be}{\begin{equation}
\addtolength{\abovedisplayskip}{\extraspaces}
\addtolength{\belowdisplayskip}{\extraspaces}
\addtolength{\abovedisplayshortskip}{\extraspace}
\addtolength{\belowdisplayshortskip}{\extraspace}}
\newcommand{\ee}{\end{equation}}

\newcommand{\ba}{\begin{eqnarray}
\addtolength{\abovedisplayskip}{\extraspaces}
\addtolength{\belowdisplayskip}{\extraspaces}
\addtolength{\abovedisplayshortskip}{\extraspace}
\addtolength{\belowdisplayshortskip}{\extraspace}}
\newcommand{\ea}{\end{eqnarray}}

\newcommand{\nonu}{\nonumber \\[.5mm]}
\newcommand{\A}{&\!\!\!}

\newcommand{\newsection}[1]{
\vspace{7mm} \pagebreak[3] \addtocounter{section}{1}
\setcounter{subsection}{0} \setcounter{footnote}{0}
\begin{center}
{\large {\bf \thesection. #1}}
\end{center}
\nopagebreak
\medskip
\nopagebreak \hspace{2mm}}

\begin{document}

\pagenumbering{arabic}

\begin{center}
{{\bf General Spherically Symmetric Non Singular Black Hole
Solutions in Teleparallel  Theory of Gravitation}}
\end{center}
\centerline{ Gamal G.L. Nashed}

\bigskip

\centerline{{\it Mathematics Department, Faculty of Science, Ain
Shams University, Cairo, Egypt }}

\bigskip
 \centerline{ e-mail:nashed@asunet.shams.edu.eg}

\hspace{2cm}
\\
\\
\\
\\
\\
\\
\\
\\

We find the most general spherically symmetric non singular black
hole solution in a special class of teleparallel theory of
gravitation. If $r$ is large enough, the general solution
coincides with the Schwarzschild solution. Whereas, if $r$ is
small, the general solution  behaves in a manner similar to that
of  de Sitter solution.  Otherwise it describes a spherically
symmetric black hole singularity free everywhere. Moreover,
 the energy associated with the general solution is calculated
 using the superpotential given by M\o ller 1978.
\newpage
\begin{center}
\newsection{\bf Introduction}
\end{center}

M\o ller modified general relativity by constructing a new field
theory in teleparallel space \cite{M1}. The aim of this theory was
to overcome the problem of energy-momentum complex that appears in
the Riemannian space \cite{M2}.  The field equations in this new
theory were derived from a Lagrangian which is not invariant under
local tetrad rotation. S$\acute{a}$ez \cite{Se} generalized M\o
ller theory into a scalar tetrad theory of gravitation.  Meyer
\cite{Me} showed that M\o ller theory is a special case of
Poincar$\acute{e}$ gauge theory \cite{HS,HNV}.

Mikhail et al. \cite{MWHL} obtained  two spherically symmetric
solutions in M\o ller's theory when the stress-energy momentum
tensor vanishing. They calculated the energy associated with those
solutions and found that in one solution the energy does not
coincide with the gravitational mass. Shirafuji et al. \cite{SNH}
extended the calculation to all the stationary asymptotic flat
space solutions with spherical symmetry, dividing them into two
classes, the first in which the components $\left(e_{a 0} \right)$
and $\left(e_{0 \alpha}\right)$ of the parallel vector fields
$\left(e_{i \mu}\right)$ tend to zero faster than $1/{\sqrt{r}}$
for large $r$ and the second, in which these components go to zero
as $1/{\sqrt{r}}$. It was found that the equality of the
gravitational and inertial masses holds only in the first class.
Mikhail et al. \cite{MWLH} also obtained a spherically symmetric
solution of M\o ller's theory starting from a tetrad which
contains three unknown functions following Mazumder and Ray
\cite{MR}. The solution contains one arbitrary function of the
radial coordinates $r$, and all the previous solutions can be
obtained from it.

Dymnikova \cite{Di} derived a static spherically symmetric
nonsingular black hole solution in orthodox general relativity
assuming a specific form of the stress-energy momentum tensor.
This solution practically coincides with the Schwarzschild
solution for large $r$,  for small $r$ it behaves like the de
Sitter solution and describes a spherically symmetric black hole
singularity free everywhere \cite{Di}. It has been proved that it
is possible to treat this specific form of the stress-energy
momentum tensor as corresponding to an r-dependent cosmological
term $\Lambda_{\mu \nu}$, varying from $\Lambda_{\mu \nu}=\Lambda
g_{\mu \nu}$ as  $r \rightarrow 0$ to $\Lambda_{\mu \nu}=0$ as $r
\rightarrow \infty$ \cite{Di1}. More recently \cite{Di2}, the
spherically symmetric nonsingular black hole has been used to
 prove that a baby universe inside a $\Lambda$ black hole can be
obtained in the case of an eternal black hole. Also it has been
shown that the probability of a quantum birth of a baby universe
can not be neglected due to the existence of an infinite number of
$\Lambda$ white hole structures. Assuming the specific form of the
vacuum stress-energy momentum tensor  \cite{Di}, using a
spherically symmetric tetrad contains three unknown functions of
radial coordinate $r$ only, which was constructed by Robertson
\cite{Ro}, the author \cite{Ga1} obtained a general solution of
this tetrad  when the stress-energy momentum tensor is not
vanishing. This solution contains one arbitrary function of the
radial coordinates $r$, and all the previous solutions could be
derived from it.

The general form of the tetrad, ${e_i}^\mu$, having spherical
symmetry was given by  Robertson \cite{Ro}. In the Cartesian form
it can be written as\footnote{In this paper Latin indices
$(i,j,...)$ represent the vector number, and Greek indices
$(\mu,\nu,...)$ represent the vector components. All indices run
from 0 to 3. The spatial part of Latin indices are denoted by
$(a,b,...)$, while that of Greek indices by $(\alpha, \beta,...).$
 In the present convention, Latin indices are never raised.}

\ba {e_0}^0 \A= \A iA, \quad {e_a}^0 = C x^a, \quad {e_0}^\alpha =
iD x^\alpha \nonu
{e_a}^\alpha \A= \A \delta_a^\alpha B + F x^a x^\alpha +
\epsilon_{a \alpha \beta} S x^\beta, \ea
 where {\it A}, {\it C}, {\it D}, {\it
B}, {\it F}, and {\it S} are functions of {\it t} and $r=(x^\alpha
x^\alpha)^{1/2}$, and the $zero^{th}$ vector ${e_0}^\mu$ has the
factor $i=\sqrt{-1}$ to preserve Lorentz signature. We consider an
asymptotically flat space-time in this paper, and impose the
boundary condition that for $r \rightarrow \infty$
 the tetrad (1) approaches the tetrad of Minkowski space-time,
$\left({e_i}^\mu\right)= {\rm diag}(i,{\delta_a}^{\alpha})$.

The aim of the present work is to find a general solution with
spherical symmetry in M\o ller's tetrad theory of gravitation when
the vacuum stress-energy momentum tensor is not vanishing and has
the same form given by  Dymnikova \cite{Di}. Moreover, the energy
of that solution is calculated. In section 2 we briefly review M\o
ller's tetrad theory of gravitation. In section 3,  we study the
general solution without the $S$-term (see 1), where the remaining
unknown functions are allowed to depend on $t$ and $r$ and  the
general solution with an arbitrary function of $t$ and $r$ is
obtained. Then we compare this general solution  with that
obtained before \cite{Ga1}, it is  found that this general tetrad
is just the $t$-independent case of our general tetrad without the
$S$-term. Also we first study the general, spherically symmetric
solution with a non-vanishing $S$-term and  a solution with one
parameter is obtained. In section 4 we calculate the energy
content of those general solutions using the superpotential
 given by M\o ller in 1978. The final section is devoted to discussion and conclusion.\\
\\
Computer algebra system Maple 6 is used in some calculations.

\newsection{M\o ller's  tetrad theory of gravitation}

In a space-time with absolute parallelism the parallel vector
fields ${e_i}^\mu$ define the nonsymmetric connection \be
{\Gamma^\lambda}_{\mu \nu} \stackrel{\rm def.}{=} {e_i}^\lambda
e_{i \ \mu, \nu}, \ee where $e_{i \mu, \nu}=\partial_\nu e_{i
\mu}$. The curvature tensor defined by ${\Gamma^\lambda}_{\mu
\nu}$ is identically vanishing, however.

M\o ller's constructed a gravitational theory based on
 this space-time. In this
theory the field variables are the 16 tetrad components
${e_i}^\mu$, from which the metric tensor is derived by \be g^{\mu
\nu} \stackrel{\rm def.}{=} {e_i}^\mu {e_i}^{\nu}. \ee
 We assume an imaginary values for the vector ${e_0}^\mu$ in
 order to have a Lorentz signature.
 We note that, associated with any tetrad field ${e_i}^\mu$ there
 is a metric field defined
 uniquely by (3), while a given metric $g^{\mu \nu}$ does not
 determine the tetrad field completely; for any local Lorentz
 transformation of the tetrads ${e_i}^\mu$ leads to a new set of
 tetrads which also satisfy (3).
  The Lagrangian ${\it L}$ is an invariant constructed from
$\gamma_{\mu \nu \rho}$ and $g^{\mu \nu}$, where $\gamma_{\mu \nu
\rho}$ is the contorsion tensor given by \be \gamma_{\mu \nu \rho}
\stackrel{\rm def.}{=} e_{i \ \mu }e_{i \nu; \ \rho}, \ee where
the semicolon denotes covariant differentiation with respect to
Christoffel symbols. The most general Lagrangian density invariant
under the parity operation is given by the form \be {\cal L}
\stackrel{\rm def.}{=} \sqrt{-g} \left( \alpha_1 \Phi^\mu
\Phi_\mu+ \alpha_2 \gamma^{\mu \nu \rho} \gamma_{\mu \nu \rho}+
\alpha_3 \gamma^{\mu \nu \rho} \gamma_{\rho \nu \mu} \right), \ee
where \be g \stackrel{\rm def.}{=} {\rm det}(g_{\mu \nu}),
 \ee
 and
$\Phi_\mu$ is the basic vector field defined by \be \Phi_\mu
\stackrel{\rm def.}{=} {\gamma^\rho}_{\mu \rho}. \ee Here
$\alpha_1, \alpha_2,$ and $\alpha_3$ are constants determined by
M\o ller
 such that the theory coincides with general relativity in the weak fields:

\be \alpha_1=-{1 \over \kappa}, \qquad \alpha_2={\lambda \over
\kappa}, \qquad \alpha_3={1 \over \kappa}(1-2\lambda), \ee where
$\kappa$ is the Einstein constant and  $\lambda$ is a free
dimensionless parameter\footnote{Throughout this paper we use the
relativistic units, $c=G=1$ and
 $\kappa=8\pi$.}. The same
choice of the parameters was also obtained by Hayashi and Nakano
\cite{HN}.

M\o ller applied the action principle to the Lagrangian density
(5) and obtained the field equation in the form \be G_{\mu \nu}
+H_{\mu \nu} = -{\kappa} T_{\mu \nu}, \ee \be F_{\mu \nu}=0, \ee
where the Einstein tensor $G_{\mu \nu}$ is defined by \be G_{\mu
\nu}=R_{\mu \nu}-{1 \over 2} g_{\mu \nu} R. \ee Here $H_{\mu \nu}$
and $F_{\mu \nu}$ are given by \be H_{\mu \nu} \stackrel{\rm
def.}{=} \lambda \left[ \gamma_{\rho \sigma \mu} {\gamma^{\rho
\sigma}}_\nu+\gamma_{\rho \sigma \mu} {\gamma_\nu}^{\rho
\sigma}+\gamma_{\rho \sigma \nu} {\gamma_\mu}^{\rho \sigma}+g_{\mu
\nu} \left( \gamma_{\rho \sigma \lambda} \gamma^{\lambda \sigma
\rho}-{1 \over 2} \gamma_{\rho \sigma \lambda} \gamma^{\rho \sigma
\lambda} \right) \right],
 \ee
and \be F_{\mu \nu} \stackrel{\rm def.}{=} \lambda \left[
\Phi_{\mu,\nu}-\Phi_{\nu,\mu} -\Phi_\rho \left({\gamma^\rho}_{\mu
\nu}-{\gamma^\rho}_{\nu \mu} \right)+ {{\gamma_{\mu
\nu}}^{\rho}}_{;\rho} \right], \ee and they are symmetric and skew
symmetric tensors, respectively.

M\o ller assumed that the energy-momentum tensor of matter fields
is symmetric. In the Hayashi-Nakano theory, however, the
energy-momentum tensor of spin-$1/2$ fundamental particles has
non-vanishing antisymmetric part arising from the effects due to
intrinsic spin, and the right-hand side of (10) does not vanish
when we take into account the possible effects of intrinsic spin.

It can be shown \cite{HS1} that the tensors, $H_{\mu \nu}$ and
 $F_{\mu \nu}$, consist of only those terms which are linear or quadratic
in the axial-vector part of the torsion tensor, $a_\mu$, defined
by \be a_\mu \stackrel{\rm def.}{=} {1 \over 3} \epsilon_{\mu \nu
\rho \sigma} \gamma^{\nu \rho \sigma}, \ee where $\epsilon_{\mu
\nu \rho \sigma}$ is defined by \be \epsilon_{\mu \nu \rho \sigma}
\stackrel{\rm def.}{=} \sqrt{-g} \delta_{\mu \nu \rho \sigma}, \ee
where $\delta_{\mu \nu \rho \sigma}$ being completely
antisymmetric and normalized as $\delta_{0123}=-1$. Therefore,
both $H_{\mu \nu}$ and $F_{\mu \nu}$ vanish if the $a_\mu$ is
vanishing. In other words, when the $a_\mu$ is found to vanish
from the antisymmetric part of the field equations, (10), the
symmetric part (9) coincides with the Einstein equation.
\newsection{Spherically symmetric solutions}

In this section we find the most general, spherically symmetric
non singular black hole solution of the form (1) in  M\o ller's
theory. The axial-vector part of the torsion tensor, $a^\mu$, is
vanishing, and the skew part of the field equation is satisfied
identically as is explained above. We discuss two cases
separately: One with $S=0$  and the other with $S \neq 0.$ \\

\underline{(i) The case without the {\it S}-term.}  In this case
the axial-vector part of the torsion tensor $a^\mu$ is identically
vanishing. Thus, when this tetrad is applied to the field
equations, the skew part (10) is automatically satisfied and the
solution of the symmetric part is the nonsingular black hole
solution given before by Dymnikova \cite{Di}. Therefore, the
solution of the form (1) with $S=0$ can be obtained from the
diagonal tetrad of the nonsingular black hole metric by a {\it
local Lorentz transformation} which keeps spherical symmetry, \be
\left(\Lambda_{k l} \right) =  \left( \matrix{ L & i H \sin\theta
\cos\phi &i H \sin\theta \sin\phi & i H \cos\theta \vspace{3mm}
\cr -i H \sin\theta \cos\phi & 1+\left(L-1 \right)\sin\theta^2
\cos\phi^2 &\left(L-1 \right)\sin\theta^2 \sin\phi \cos\phi
&\left(L-1 \right)\sin\theta \cos\theta \cos\phi \vspace{3mm} \cr
-i H \sin\theta \sin\phi &\left(L-1 \right) \sin\theta^2 \sin\phi
\cos\phi &1+\left(L-1 \right)\sin\theta^2 \sin\phi^2 &\left(L-1
\right)\sin\theta \cos\theta \sin\phi \vspace{3mm} \cr -i H
\cos\theta &\left(L-1 \right)\sin\theta \cos\theta \cos\phi
&\left(L-1 \right)\sin\theta \cos\theta \sin\phi  &1+\left(L-1
\right)\cos\theta^2 \cr}\right), \ee where ${\it H}$ is an
arbitrary function of {\it t} and $\hat{R}$ and
\[L=\sqrt{H^2+1}.\]
 Namely, we see that \be {e_i}^\mu= \Lambda_{i
l} {e_l}^{(\small 0) \mu} \ee is the most general, spherically
symmetric solution without the {\it S}-term. Here ${e_l}^{ (\small
0) \mu}$ is the diagonal tetrad in the spherical polar coordinates
given by \cite{Ga}

\be \left({e_l}^{ (\small 0) \mu} \right)= \left( \matrix{
\displaystyle{i \over \hat{X}} &0 & 0 & 0 \vspace{3mm} \cr 0 &
\hat{X} \sin\theta \cos\phi  & \displaystyle{\cos\theta \cos\phi
\over \hat{R}}
 & -\displaystyle{ \sin\phi  \over \hat{R} \sin\theta} \vspace{3mm} \cr
0 & \hat{X} \sin\theta \sin\phi  & \displaystyle{\cos\theta
\sin\phi \over \hat{R}}
 & \displaystyle{\cos\phi \over \hat{R} \sin\theta} \vspace{3mm} \cr
0 & \hat{X} \cos\theta & -\displaystyle{\sin\theta \over \hat{R}}
& 0 \cr } \right), \ee where \[\hat{X}=\sqrt{1- \displaystyle{2m
\over \hat{R} } \left(1-e^{-\hat{R}^3/{r_1}^3}\right)}\] and
$\hat{R}$ is defined as $\hat{R}={r/B}$ \cite{Ga}. The explicit
form of the ${e_i}^\mu$ is then given by \newpage \be
\left({e_i}^\mu \right) = \left( \matrix{ \displaystyle{i L \over
\hat{X}} &i H \hat{X} &0 & 0 \vspace{3mm} \cr \displaystyle{H
\sin\theta \cos\phi \over \hat{X} } & L \hat{X} \sin\theta
\cos\phi & \displaystyle{ \cos\theta \cos\phi \over \hat{R}} &
-\displaystyle{\sin\phi \over \hat{R} \sin\theta} \vspace{3mm} \cr
\displaystyle{H \sin\theta \sin\phi \over \hat{X}} & L \hat{X}
\sin\theta \sin\phi & \displaystyle{ \cos\theta \sin\phi \over
\hat{R}} & \displaystyle{ \cos\phi \over \hat{R} \sin\theta}
\vspace{3mm} \cr \displaystyle{H \cos\theta \over \hat{X}} & L
\hat{X} \cos\theta &-\displaystyle{\sin\theta \over \hat{R}} & 0
\cr}\right), \ee

where $r_1$ is defined by \ba {r_1}^3 \A=\A r_g {r_0}^2,\nonu
 r_g\A=\A 2m,\nonu
{r_0}^2 \A=\A {3 \over 8\pi\epsilon_0}.
 \ea

 If we apply the tetrad (19) to the symmetric part of the field
equation (9), the right hand side takes the form
 \ba
  {T_0}^0={T_1}^1 \A=\A
\epsilon_0 e^{-\hat {R}^3/{r_1}^3},\nonu
{T_2}^2={T_3}^3 \A=\A \epsilon_0 e^{-\hat {R}^3/{r_1}^3}
\left(1-{3 \hat {R}^3 \over 2{r_1}^3} \right). \ea The metric
associated with the tetrad (19) is given by \be ds^2=-\xi
dt^2+{d\hat {R}^2 \over \xi}+\hat {R}^2 d\Omega^2, \ee where \be
\xi=\hat{X}^2, \quad  and \qquad {d\Omega^2=d\theta^2+\sin^2\theta
d\phi^2},\ee which is the spherically symmetric nonsingular black
hole solution \cite{Di}.

\underline{(ii) The case with non-vanishing {\it S}-term.}  We
start with the tetrad of (1) with the six unknown functions of
{\it t} and {\it r}. In order to study the condition that the
$a^\mu$ vanishes it is convenient to start from the general
expression for the covariant components of the tetrad, \ba e_{0
\scriptstyle{0}} \A= \A i \check{A}, \quad e_{a \scriptstyle{0}}=
\check{C} x^a, \quad  e_{0 \scriptstyle{\alpha}}= i \check{D}
x^\alpha \nonu
e_{a \scriptstyle{\alpha}} \A= \A \delta_{a \alpha}
\check{B}+\check{F} x^a x^\alpha + \epsilon_{a \alpha \beta}
\check{S} x^\beta, \ea where the six unknown functions,
$\check{A}$, $\check{C}$, $\check{D}$, $\check{B}$, $\check{F}$,
$\check{S}$, are connected with the six unknown functions of (1).
We can assume without loss of generality that the two functions,
$\check{D}$ and $\check{F}$, are vanishing by making use of the
freedom to redefine ${\it t}$ and ${\it r}$ \cite{HS1}.  We then
transform the tetrad (24) to the spherical polar coordinates
$(r,\theta,\phi,t)$ to take the form \be \left(e_{i
\scriptstyle{\mu}} \right)= \left( \matrix{ iA & 0 & 0 & 0
\vspace{3mm} \cr r \check{C} \sin\theta \cos\phi & \check{B}
\sin\theta \cos\phi & r \check{B}\cos\theta
\cos\phi+r^2\check{S}\sin\phi
 & -r \check{B} \sin\theta \sin\phi+r^2\check{S} \sin\theta \cos\theta \cos\phi \vspace{3mm} \cr
 r \check{C} \sin\theta \sin\phi & \check{B}
\sin\theta \sin\phi &r \check{B}\cos\theta
\sin\phi-r^2\check{S}\cos\phi  & r \check{B} \sin\theta
\cos\phi+r^2\check{S} \sin\theta \cos\theta \sin\phi \vspace{3mm}
\cr r \check{C} \cos\theta  & \check{B} \cos\theta & -r \check{B}
\sin\theta  & -r^2\check{S}\sin^2\theta \cr } \right). \ee
 Then the condition that the axial vector part $a^{\mu}$ vanishes is
to be \cite{SNH}.

\be
 0 = \sqrt{(-g)} a^\mu = \left\{
\matrix{ & 3 \check{B} \check{S}+ r(\check{B} \check{S}'
-\check{B}' \check{S}), \quad \mu=0, \hfill\cr &
 2 \check{C}\check{S}+ (\check{\dot{S}}\check{B} -\check{S}\check{\dot{B}}),
 \qquad \mu=1 \hfill\cr }\right. \ee with $\check{S}'= {d
\check{S}/dr}$  and $\check{\dot{S}}= {d \check{S}/dt}$. This
condition can be solved to give \be
 \check{C}=0, \quad \check{S}={{\eta} \over r^3} \check{B},
\ee where $\eta$ is a constant with dimension of $(length)^2$.

The symmetric part of the field equations now coincides with the
Einstein equation. The metric tensor formed of the tetrad (25)
with (27) is not of the non singular black hole solution form.
Taking the new radial coordinate \be R =r \check{B}
\sqrt{1+{\eta^2 \over r^4}}, \ee then the metric tensor takes the
well-known non singular black hole solution form \cite{Di}.

Applying the transformation (28) to the tetrad (25) with (27), we
finally obtain the general, spherically symmetric non singular
black hole solution with non-vanishing {\it S}-term: The
non-vanishing, covariant components of the tetrad are given by \ba
e_{0 \scriptstyle{0}} \A= \A i X \nonu
 e_{1 \scriptstyle{1}} \A= \A
 \displaystyle{\sin\theta \cos\phi \over X} \nonu
e_{1 \scriptstyle{2}} \A= \A \displaystyle{R \over Y} \left(4 \
\eta \ f^2(R) \ \sin\phi+4 \ f^4(R) \ \cos\theta \cos\phi - \eta^2
\ \cos\theta \cos\phi \right)  \nonu
e_{1 \scriptstyle{3}} \A= \A \displaystyle{R \over Y} \left(4 \
\eta \ f^2(R) \  \cos\theta\cos\phi -4 \ f^4(R) \ \sin\phi+ \eta^2
\ \sin\phi \right)\sin\theta \nonu
e_{2 \scriptstyle{1}} \A= \A
 \displaystyle{\sin\theta \sin\phi \over X} \nonu
e_{2 \scriptstyle{2}} \A= \A \displaystyle{-R \over Y} \left(4\
\eta \ f^2(R) \ \cos\phi -4 \ f^4(R) \ \cos\theta \sin\phi +
\eta^2\ \ \cos\theta \sin\phi \right)  \nonu
e_{2 \scriptstyle{3}} \A= \A \displaystyle{R \over Y} \left(4 \
\eta \ f^2(R) \ \cos\theta\sin\phi+4 \ f^4(R) \ \cos\phi- \eta^2 \
\cos\phi \right)\sin\theta \nonu
e_{3 \scriptstyle{1}} \A= \A
 \displaystyle{\cos\theta \over X} \nonu
e_{3 \scriptstyle{2}} \A= \A \displaystyle{-R \over Y}
 \left(4f^4(R)-\eta^2\right) \sin\theta  \nonu
e_{3 \scriptstyle{3}} \A= \A \displaystyle{-R \over Y} 4 \ \eta
f^2(R) \ \sin^2\theta, \ea where \[X=\sqrt{1-\displaystyle{2m
\over R}\left(1-e^{-R^3/{r_1}^3} \right)},\qquad
Y=4f^4(R)+\eta^2\]  and $f(R)$ is given by \be f(R)=\large {e^{
\int dR/(R X)}},\ee

The tetrad (29) when applies to the symmetric part of field
equation (9), the right hand side takes the form
 \ba
  {T_0}^0={T_1}^1 \A=\A
\epsilon_0 e^{-R^3/{r_1}^3},\nonu
{T_2}^2={T_3}^3 \A=\A \epsilon_0 e^{-R^3/{r_1}^3} \left(1-{3R^3
\over 2{r_1}^3} \right). \ea The metric associated with the tetrad
(29) is given by \be ds^2=-\xi_1 dt^2+{dR^2 \over \xi_1}+R^2
d\Omega^2, \ee where \be \xi_1={X}^2, \ee
 which is the spherically symmetric
  nonsingular black hole solution \cite{Di}.

 It is clear that if $\eta$ and $H(\hat{R},t)$ are equal to zero the
two classes of solutions given by \\ (19) and (29) coincide with
each other, and reduce to the solution given before \cite{Ga}.
 Furthermore, if the exponential term is equal to zero then the
 two solutions
coincides with each other and give the Schwarzschild solution.

The solutions (19) and (29) are the exact solutions of the M\o
ller's field equations. They  practically coincide with the
Schwarzschild solution for $R>>r_1$ or $\hat{R}>>r_1$ and  behave
like the de Sitter solution, for $R<<r_1$ or $\hat{R}<<r_1$.

As is clear from (21), the spherically symmetric stress-energy
momentum  tensor is really anisotropic. The difference between the
principle pressures \be {T_k}^k=-p_k, \ee correspond to the well
known anisotropic character of evolution of the space-time inside
a black hole undergoing a spherically symmetric gravitational
collapse \cite{ZN}. For $R<<r_1$ isotropization occurs and the
stress-energy momentum tensor takes the isotropic form \be
T_{\alpha \beta}=\epsilon g_{\alpha \beta}. \ee When $R\rightarrow
0$ the energy density tends to $\epsilon_0$. For $R>>r_1$ all the
components of the stress-energy momentum tensor tend to zero very
rapidly. The same properties satisfied for the stress-energy
momentum tensor (31).

Now let us compare the solution (19) with that given before
  \cite{Ga1}. Started from a spherically symmetric tetrad
with three unknown functions of the radial coordinate $r$ only,
which is given in the spherical polar coordinates by \be
\left({e_i}^\mu \right)= \left( \matrix{ iA & iDr & 0 & 0
\vspace{3mm} \cr 0 & B \sin\theta \cos\phi & \displaystyle{B \over
r}\cos\theta \cos\phi
 & -\displaystyle{B \sin\phi \over r \sin\theta} \vspace{3mm} \cr
0 & B \sin\theta \sin\phi & \displaystyle{B \over r}\cos\theta
\sin\phi
 & \displaystyle{B \cos\phi \over r \sin\theta} \vspace{3mm} \cr
0 & B \cos\theta & -\displaystyle{B \over r}\sin\theta  & 0 \cr }
\right). \ee Applying the tetrad (36)  to the field equations, (9)
and (10), we obtained a general solution of the form \ba {\cal A}
(\hat {R}) \A=\A \displaystyle{1 \over 1-\hat{R} {\cal B}'},\nonu
{\cal D}(\hat {R}) \A=\A\displaystyle{1 \over 1-\hat{R} {\cal B}'}
\sqrt{\displaystyle{2m \over \hat{R}^3} \left(1-{\huge
e^{(-\hat{R}^3/{r_1}^3)}} \right)+ \displaystyle{{\cal B}' \over
\hat{R}} \left( \hat{R} {\cal B}' -2 \right)}, \ea where ${\cal
A}(\hat {R})$, \quad ${\cal B}(\hat {R})$ \quad and \quad ${\cal
D}(\hat {R})$ are connected  to $A$, $B$ and  $D$ through the
transformation \\ $\hat{R}=r/B$, \quad ${\cal
B}'=\displaystyle{d{\cal B}(\hat{R}) \over d \hat{R}}$, \quad and
${\it \cal{B}}$ is an arbitrary function of ${\it \hat{R}}$. The
tetrad (36) can further be expressed in the form \cite{Ga1} \be
\left({e_i}^\mu \right)= \left( \matrix{ \displaystyle{i{\cal A}
\over 1-{\cal D}^2 \hat{R}^2} & i{\cal D} \hat{R}(1-\hat{R}{\cal
B}') & 0 & 0 \vspace{3mm} \cr \displaystyle{{\cal A} {\cal D}
\hat{R} \sin\theta \cos\phi \over 1-{\cal D}^2\hat{R}^2}
&(1-\hat{R} {\cal B}') \sin\theta \cos\phi &
\displaystyle{\cos\theta \cos\phi \over \hat{R}}
 & -\displaystyle{\sin\phi \over \hat{R} \sin\theta} \vspace{3mm} \cr
\displaystyle{{\cal A} {\cal D} \hat{R}  \sin\theta \sin\phi \over
1-{\cal D}^2 \hat{R}^2} & (1-\hat{R}{\cal B}') \sin\theta \sin\phi
&
 \displaystyle{\cos\theta \sin\phi \over \hat{R}}
 & \displaystyle{\cos\phi \over \hat{R} \sin\theta} \vspace{3mm} \cr
\displaystyle{{\cal A} {\cal D} \hat{R} \cos\theta  \over 1-{\cal
D}^2\hat{R}^2} & (1-\hat{R}{\cal B}') \cos\theta &
\displaystyle{-\sin\theta \over \hat{R}} & 0 \cr } \right). \ee
Thus, it is easy to verify that this tetrad can be obtained from
(19) by choosing the function $H$ as \be H={\left[\hat{R}^2 {\cal
B}'^2-2\hat{R} {\cal B}'+\displaystyle{2m \over
\hat{R}}\left(1-{\huge e^{(-\hat{R}^3/{r_1}^3)}} \right)
\right]^{1/2} \over \sqrt{1-\displaystyle{2m \over
\hat{R}}\left(1-e^{-\hat{R}^3/{r_1}^3}\right)}} \ee
\newsection{The Energy Associated with each Solution}

The superpotential of the M\o ller's tetrad theory of gravitation
 is given by Mikhail et al. \cite{MWHL} as
\be {{\cal U}_\mu}^{\nu \lambda} ={(-g)^{1/2} \over 2 \kappa}
{P_{\chi \rho \sigma}}^{\tau \nu \lambda} \left[\Phi^\rho
g^{\sigma \chi} g_{\mu \tau}
 -\lambda g_{\tau \mu} \gamma^{\chi \rho \sigma}
-(1-2 \lambda) g_{\tau \mu} \gamma^{\sigma \rho \chi}\right], \ee
where ${P_{\chi \rho \sigma}}^{\tau \nu \lambda}$ is \be {P_{\chi
\rho \sigma}}^{\tau \nu \lambda} \stackrel{\rm def.}{=}
{{\delta}_\chi}^\tau {g_{\rho \sigma}}^{\nu \lambda}+
{{\delta}_\rho}^\tau {g_{\sigma \chi}}^{\nu \lambda}-
{{\delta}_\sigma}^\tau {g_{\chi \rho}}^{\nu \lambda} \ee with
${g_{\rho \sigma}}^{\nu \lambda}$ being a tensor defined by \be
{g_{\rho \sigma}}^{\nu \lambda} \stackrel{\rm def.}{=}
{\delta_\rho}^\nu {\delta_\sigma}^\lambda- {\delta_\sigma}^\nu
{\delta_\rho}^\lambda. \ee The energy is expressed by the surface
integral \cite{Mo2} \be E=\lim_{r \rightarrow
\infty}\int_{r=constant} {{\cal U}_0}^{0 \alpha} n_\alpha dS, \ee
where $n_\alpha$ is the unit 3-vector normal to the surface
element ${\it dS}$.

Now we are in a position to calculate the energy associated with
 the two solutions (19) and (29) using the superpotential (40). As is
 clear from (43), the only component which contributes to the energy is
  ${{\cal U}_0}^{0
 \alpha}$. Thus substituting from the solution (19) into
 (40) we obtain the following non-vanishing value

 \be
{{\cal U}_0}^{0 \alpha}={2 \hat{X}  x^\alpha \over \kappa
\hat{R}}\left(L-\hat{X} \right).
 \ee
 Substituting from (44) into
(43) we get \be E(\hat {R})=\hat{X} \hat{R} \left(L-\hat{X}
\right). \ee As is clear from (45) that the energy depends on the
arbitrary function $H$. If this arbitrary function takes the value
(39) then \be E(\hat {R})=2m \left(1-e^{-\hat
{R}^3/{r_1}^3}\right)-\hat {R}^2 {\cal B}', \ee which is the same
as that obtained before \cite{Ga1}.

 Now let us turn our attention to the solution (29). Calculating
the necessary components of the superpotential, we get

 \be
{{\cal U}_0}^{0 \alpha}={4X x^\alpha \over \kappa
R}\left({(4f(R)^4-\eta^2) \over Y}-X \right). \ee Substituting
from (47) into (43) we get \be E(R)=X R \left({(4f(R)^4-\eta^2)
\over Y}-X \right), \ee where $f(R)$ is given by (30). As is clear
from (48) that if the constant $\eta=0$, then \be
E(R)=m(1-e^{-R^3/{r_1}^3}), \ee which is the same as that given
before  \cite{Ga}, Yang \cite{Yi} and Radinschi \cite{Ra}. This is
a very satisfactory result since the two tetrads (29), (36)
coincide when $\eta=0$ and $D=0$, and accordingly their energies
must be identical. We can also see from (48) that if $R\rightarrow
0$, then $E(R)\rightarrow 0$ and form (49) if $R\rightarrow
\infty$ then $E(R)\rightarrow m$. Thus the total energy will
coincide with that of the Schwarzschild solution.
\newpage
\newsection{Main results and Discussion}
We have studied the general spherically symmetric nonsingular
black hole solutions in teleparallel theory of gravitation.
Tetrads with spherical symmetry are classified into two groups. \vspace{.1cm}\\
i) The groups in which the S-term is vanishing, and therefore, the
axial vector part $a^\mu$ is identically vanishing. Accordingly we
obtain a family of solutions with arbitrary function of t and
$\hat {R}$, i.e., $H(t,\hat{R})$. A relation with the solution
given before  \cite{Ga1}  had been established by assuming a
specific form of the arbitrary function $H(t,\hat{R})$ as given by
(39).\vspace{.1cm}\\ ii) The groups which possess the S-term,
namely, the term $S(t,r)\epsilon_{a \alpha \beta}x^\beta$. When
the S-term is non-vanishing, we obtain the condition (27) which is
the condition for vanishing the axial vector part, i.e. $a^\mu=0$.
In this case we obtain a family of solutions with a constant
parameter, i.e., $\eta$. For comparison if this constant is set
equal to zero then the tetrad (29) will reduce to the diagonal
form of (36).

Pereira et al. \cite{PVZ} studied the teleparallel  version of the
exact solutions ( Schwarzschild and Kerr solutions) of general
relativity. As is well known that the torsion tensor can be
decomposed  into three parts tracless part, vector part and the
axial vector part \cite{HS1}. Using the {\it weak-field limit}
they \cite{PVZ} show that the vector and the tracless parts of the
Schwarzschild torsion combine  to yield the Newtonian force. By
considering the {\it slow rotation and weak-field approximation}
they show that for the Kerr solution the axial vector part is
nothing but the gravitomagnetic components of the gravitational
field and is therefore the responsible for the Lense Thirring
effect.

We also calculated the energy associated with the two solutions
(19) and (29). Concerning the energy associated with the solution
  without the S-term, it is found that the energy depends on the
arbitrary function $H(t,\hat{R})$. If the function $H(t,\hat{R})$
takes the value (39), then the energy coincides with that obtained
before  \cite{Ga1} and if $H(t,\hat{R})=0$, then the energy
coincides with that obtained  by  Yang \cite{Yi} and Radinschi
\cite{Ra}

As for the energy associated with the solution when the S-term is
non-vanishing we obtain the expression (48). It is clear from this
expression that the energy depends on the parameter $\eta$ and if
this parameter equals zero then the tetrad (29) will coincide with
the diagonal form of the tetrad (36) and the associated energy
will be the same as that obtained before  \cite{Ga}. It is clear
from (48), that if $R\rightarrow 0$, then $E(R)\rightarrow 0$ and
from (49)  if $R\rightarrow \infty$ then $E(R)\rightarrow m$. Thus
the total energy will coincide with that of the Schwarzschild
solution, since the metric associated with the tetrad (29) in this
case gives the Schwarzschild metric.

It is readily  seen from equations (46) and (49) that the factors
in front of the gravitational mass m are different. This is
because that when $H(t,\hat{R})=0$,
 the components $\left(e_{a 0} \right)$ and $\left(e_{0
\alpha}\right)$ of the parallel vector fields (19) tend to zero as
$1/{\sqrt{\hat{R}}}$ and in that case the gravitational mass will
not coincide with the inertial mass even in the simple case of
vacuum, i.e., $\epsilon_0=0$ \cite{MWHL, SNH, SNK}. In contrast to
the tetrad (29) when $\eta=0$, the components $\left(e_{a 0}
\right)$ and $\left(e_{0 \alpha}\right)$
 tend to zero faster than  $1/{\sqrt{R}}$ and this case is
 satisfactory since the gravitational mass equals to the inertial
 mass \cite{MWHL, SNH}.

 In the case of the ADM mass we obtain the same result as we can
 see from (49)
 \[ M_{ADM}=E(R)_{R\rightarrow \infty}=m. \]
But for equation (46) the result will depend on the form of the
arbitrary function ${\cal B}$.

We calculate the energy using expression (43) mainly given by M\o
ller \cite{M1}. Blagojevi$e'$ and Vasili$e'$ \cite{BV} presented
an investigation of the connection between the asymptotic
Poincar$e'$ symmetry of space-time and the related conservation
laws of energy-momentum and angular momentum in teleparallel
theory of gravity. They \cite{BV} derived the generators of the
global Poincar$e'$ symmetry in the asymptotic region from the
related gauge generators. They \cite{BV} concluded that the
Poincar$e'$ generators have to be improved by adding certain
surface terms, which represent the values of energy-momentum and
angular momentum of the gravitating system. These results for
energy-momentum and angular momentum are valid for general
teleparallel theory of gravity.

 Chang et al. \cite{CNC} argued that every energy-momentum
complex is associated with a legitimate Hamiltonian boundary term,
and, because of this the energy-momentum complexes are quasi-local
and acceptable. Each is the energy-momentum density for some
physical situation. This Hamiltonian approach to quasi-local
energy-momentum rehabilitates of the energy-momentum complexes.

A summary of the main results is given in the table below. The
general solution of spherically symmetric nonsingular black hole
is classified into two groups according to whether or not the
space-space components ${e_a}^\alpha$ have the term $S\epsilon_{a
\alpha \beta}x^\beta$ (referred to as the $S$-term for short). The
general solution without the $S$-term has an arbitrary function of
$t$ and $r$. The general solution with the $S$-term has a constant
parameter.

\begin {center}
\begin{tabular}{|c|c|c|c|} \hline
  & \multicolumn{2} {|c|} { Field equation} & Energy \\
\cline{2-3}  & skew part & Symmetric part& \\ \hline Tetrad &
Satisfied &  Nonsingular black&$\hat{X} \hat{R}\left( L -\hat{X}
\right)$
 \\
without the $S$-term &  identically & hole solution &

\\ \hline
Tetrad & Gives &  Nonsingular black &$X R
\left(\displaystyle{(4f(R)^4-\eta^2)
 \over Y}-X \right)$
 \\
with $S$-term & $a^\mu=0$ & hole solution&  \\ \hline
\end{tabular}
\end {center}

\bigskip
\bigskip
\centerline{\Large{\bf Acknowledgements}}

The author would like to thank Professor I.F.I. Mikhail; Ain Shams
University, for his  stimulating discussions.

\end{document}